\documentclass[graphicx,superscriptaddress,prb,reprint,longbibliography]{revtex4-1}

\usepackage[version=3]{mhchem} 
\usepackage[T1]{fontenc}
\usepackage{siunitx}
\usepackage{graphicx}
\usepackage{eufrak}

\usepackage[normalem]{ulem}
\useunder{\uline}{\ul}{}
\begin{document}
\author{Robin J. Dolleman}
\email{dolleman@physik.rwth-aachen.de}
\affiliation{Kavli Institute of Nanoscience, Delft University of Technology, Lorentzweg 1, 2628 CJ, Delft, The Netherlands}
\affiliation{Present adress: Second Institute of Physics, RWTH Aachen University, 52074, Aachen, Germany}
\author{Allard J. Katan}
\affiliation{Kavli Institute of Nanoscience, Delft University of Technology, Lorentzweg 1, 2628 CJ, Delft, The Netherlands}
\author{Herre S. J. van der Zant}
\affiliation{Kavli Institute of Nanoscience, Delft University of Technology, Lorentzweg 1, 2628 CJ, Delft, The Netherlands}
\author{Peter G. Steeneken}
\email{P.G.Steeneken@tudelft.nl}
\affiliation{Kavli Institute of Nanoscience, Delft University of Technology, Lorentzweg 1, 2628 CJ, Delft, The Netherlands}
\affiliation{Department of Precision and Microsystems Engineering, Delft University of Technology, Mekelweg 2, 2628 CD, Delft, The Netherlands}
\title{Semi-permeability of graphene nanodrums in sucrose solution}

\begin{abstract}
Semi-permeable membranes are important elements in water purification and energy generation applications, for which the atomic thickness and strength of graphene can enhance efficiency and permeation rate while maintaining good selectivity. Here, we show that an osmotic pressure difference forms across a suspended graphene membrane as a response to a sucrose concentration difference, providing evidence for its semi-permeability.
This osmotic pressure difference is detected via the deflection of the graphene membrane that is measured by atomic force microscopy. Using this technique, the time dependence of this deflection allows us to measure the water permeation rate of a single 3.4 $\mu$m diameter graphene membrane. Its value is close to the expected value of a single nanopore in graphene. The method thus allows one to experimentally study the semi-permeability of graphene membranes at the microscale when the leakage rate is miniscule. It can therefore find use in the development of graphene membranes for filtration, and can enable sensors that measure the concentration and composition of solutions.
\end{abstract}
\maketitle

\section{Introduction}
Semi-permeable membranes are of essential for filtration and separation in the chemical, food and pharmaceutical industry \cite{basile2011advanced}. Membrane technology plays a growing role in the transition to a sustainable society, for example in energy storage applications like fuel cells and batteries \cite{che1998carbon,tan2020hydrophilic},  in energy generation in osmotic power plants \cite{logan2012membrane} and in water purification \cite{ISI:000254117400037}. Recent developments in nanotechnology allow for fabrication of sub-nanometer pores to improve the selectivity of semi-permeable membranes, enabling new applications \cite{epsztein2020towards}. 
A small thickness is beneficial for separation membranes, because it allows high flow rates at small power consumption \cite{cheryan1998ultrafiltration}. Therefore graphene, consisting of a single layer of carbon atoms bonded in a hexagonal lattice \cite{geim2007rise}, is a promising candidate for future semi-permeable membranes with ultimate performance. Besides its small thickness, also the chemical stability \cite{EFTEKHARI20171} and mechanical strength of graphene \cite{lee2008measurement,zandiatashbar2014effect,wang2017single} are advantageous. These properties have attracted considerable attention for studies into graphene-based water purification \cite{o2012selective,surwade2015water,o2015nanofiltration,cohen2012water,humplik2011nanostructured,cohen2015nanoporous,rollings2016ion,suk2014ion,cohen2014water,suk2010water,boretti2018outlook}, gas separation \cite{boutilier2014implications,jiang2009porous,kim2013selective,celebi2014ultimate} and gas sensing \cite{koenig2012selective,dolleman2016graphene,roslon2020graphene}.
Water permeation studies on graphene often require macroscopic setups to apply an hydraulic pressure and measure the permeation rate. Therefore, the permeation occurs over a large area, making it difficult to verify molecular dynamics simulations performed at the nanoscale. Moreover, a single defect in a large membrane can dominate the permeation rate and thus significantly impact the total membrane performance. To further understand water permeability of graphene at the single pore level, it is therefore of interest to develop experimental techniques to measure the permeability on a microscopic level.

Here, we demonstrate the semi-permeability of single graphene nanodrums, which are graphene membranes suspended over circular cavities in a substrate, with an area of {9 $\mu$m$^2$}. The semi-permeability is demonstrated by monitoring their deflection using a liquid atomic force microscopy (AFM) technique. It is observed that suspended graphene membranes sealing a cavity are deflected when a sucrose concentration difference is applied across the membrane. This deflection is attributed to the osmotic pressure generated on the membrane, which pushes the water out of the cavity and reduces its volume. The time-dependence of this deflection allows determination of the water permeation rate of the graphene drum and this continues until an equilibrium between the osmotic force and membrane force is obtained. Thus facilitating the water permeability of graphene to be studied at microscopic scales where the volumetric flux is miniscule. 

\section{Experimental methods}
\begin{figure}[t]
\includegraphics{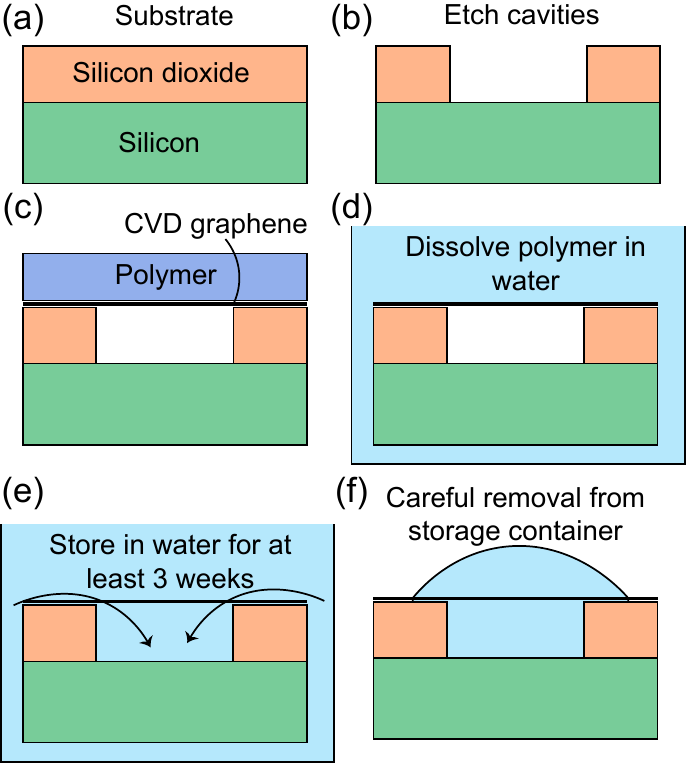}
\caption{Illustration of the sample fabrication. (a) Fabrication starts on a 19x19 mm$^2$ silicon die with 285 nm thermally grown silicon dioxide. (b) Cavities are etched in the silicon dioxide layer using reactive ion etching. (c) Single--layer CVD graphene is transferred using a water soluble polymer. (d) The polymer is dissolved in water. (e) The sample is kept in water in a storage container for at least 3 weeks to let gases permeate from the cavitity and water permeate in. (f) Before the experiment the sample is carefully removed from the storage container, keeping a water droplet on the sample to ensure the cavities stay submerged. \label{fig:preparation}}
\end{figure}
The steps taken for sample preparation are shown in Fig. \ref{fig:preparation}. Fabrication starts from a silicon chip with a layer of 285 nm of thermally grown silicon dioxide (Fig. \ref{fig:preparation}a). Circular cavities of 3.4 $\mu$m diameter are patterned using electron beam lithography and etched to a depth of 285 nm in the oxide layer using reactive ion etching (Fig. \ref{fig:preparation}b). A sheet of single-layer graphene grown by chemical vapor deposition (CVD) is transferred on top of the chip and suspended over the cavities using a water dissolvable transfer polymer (Fig. \ref{fig:preparation}c). The transfer polymer is dissolved in water, after which the sample is kept wet during the rest of the fabrication and experiment (Fig. \ref{fig:preparation}d). To allow water to permeate into the cavities, and to let the gas in the cavity permeate out and dissolve in the water, the sample is stored in a container with deionized (DI) water for more than three weeks before the experiment is started (Fig. \ref{fig:preparation}e). By studying the samples in a liquid cell AFM, this sample preparation results in cavities filled with water in two out of the four samples; on the other two samples only broken drums are found.

\begin{figure}[t]
\includegraphics{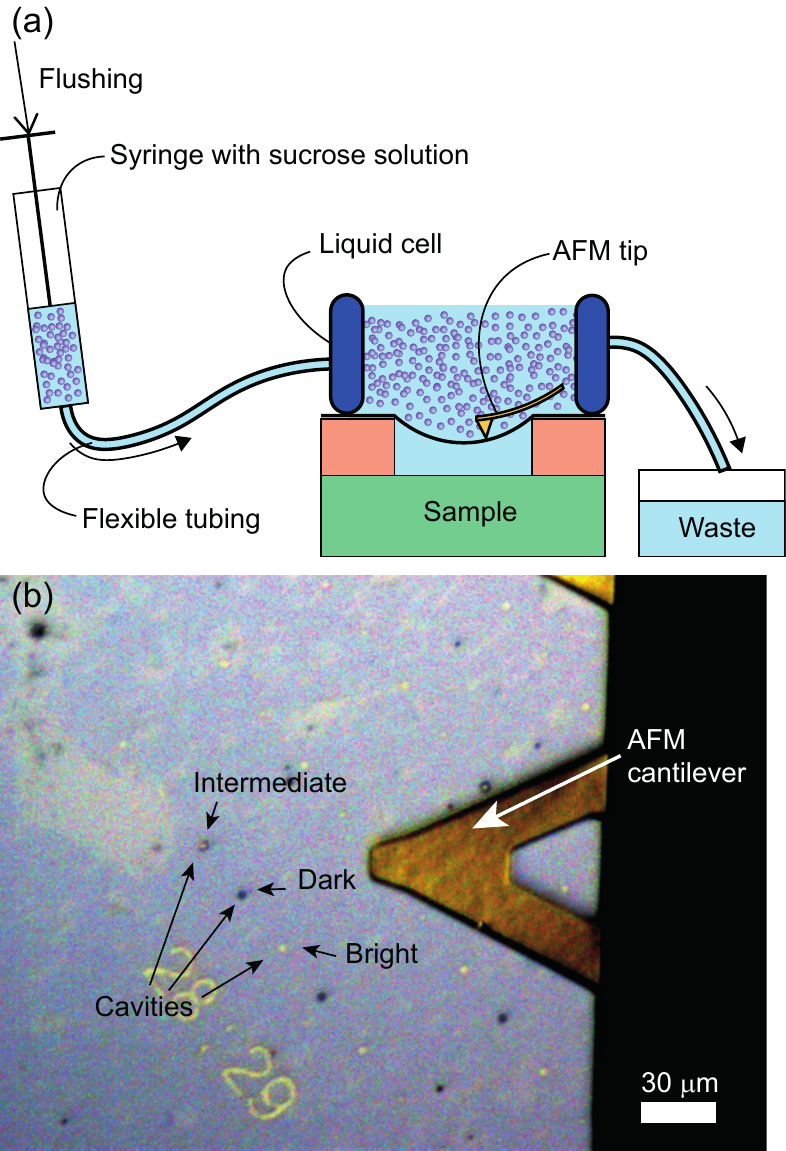}
\caption{(a) Schematic of the measurement setup to detect the deflection due to osmotic pressure across the graphene membrane. (b) Optical image taken of the AFM tip on the sample during the experiment. \label{fig:setup}}
\end{figure}
 The deflection of the membranes is studied using atomic force microscopy in a liquid cell shown in Fig. \ref{fig:setup}a. Since the surface tension of water may break the graphene nanodrum, the cavities have to remain submerged. Therefore, the sample is removed from the container in such a way that a droplet of water remains on the center part of the chip (Fig. \ref{fig:preparation}f). The sample is then moved to an AFM with a flexible silicon-rubber liquid cell. The 19 by 19 mm$^2$ chip forms the bottom of the liquid cell, while the rubber encapsulation of the liquid cell ensures that the chip remains fully immersed in water during the experiment. AFM imaging is performed on a Bruker Dimension Icon AFM system operating in tapping mode. The cantilevers used are Bruker ScanAsyst Fluid, with a nominal stiffness of 0.7 N/m. The AFM scans over the surface until a suspended graphene drum is found that fully covers the cavity. From the optical image in the AFM we observe that the cavities have generally three different optical contrasts (Fig. \ref{fig:setup}b); some appear bright, others dark and some show intermediate contrast. All the drums used for the experiment show an intermediate optical contrast. From results of initial AFM tests it is hypothesized that the brightest and darkest cavities correspond to cavities with broken graphene and air-filled cavities sealed with graphene, respectively, but this was not studied in more detail. 

 The liquid cell has two flexible tubes connected to it (Fig. \ref{fig:setup}a), which can be used to flush the cell with a solution. A syringe with a sucrose solution of a well-known concentration is connected to one of the tubes and the system is flushed with 2 mL of the solution, which is a volume much larger than the volume of the liquid cell and the tubing (in the order of 100 $\mu$L), such that the concentration outside the cavity is equal to the concentration in the syringe. Since the concentration in the cavity remains zero, a well-known concentration difference is applied across the membrane. By flushing the liquid cell slowly and carefully to minimize mechanical disturbances, the AFM can remain in contact with the substrate and continue scanning without having to retract the AFM tip from the surface. After flushing, the AFM continues to scan the drum in order to measure the height of the membrane as a function of time. 
 
\section{Results}
\begin{figure*}
\includegraphics{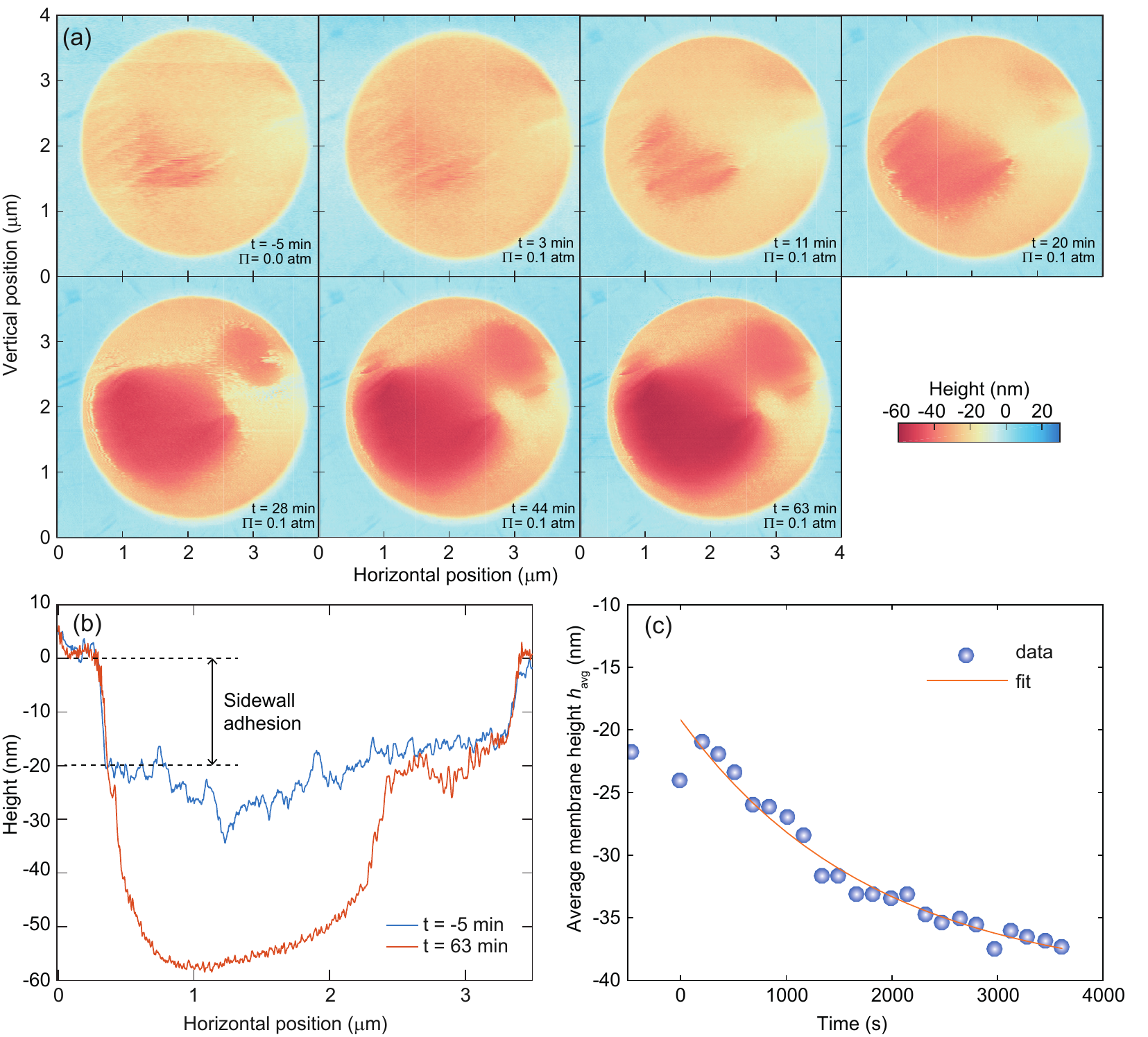}
\caption{Time-dependent deflection of a 3.4-micron diameter graphene drum subjected to 0.1 atm of osmotic pressure. (a) Height scans of the drum at different times. Flushing with a 1.6 g/L sucrose solution with osmotic pressure $\Pi = 0.1$ atm starts at $t = -2$ min and ends at $t= 0$ min. (b) Horizontal cross section along the center of the drum (at a vertical position of 2 $\mu$m) at times $t = -5$ min and $t = 63$ min. (c) Average height of the suspended membrane as a function of time. From fitting Eq. \ref{eq:timedep} to the experimental data, we find $n_{\mathrm{mem}} = 0.13$ N/m and $\tau = \num{1.8e3}$ s.  \label{fig:afmdefl}} 
\end{figure*}
Figure \ref{fig:afmdefl}a shows the height map of a 3.4-micron diameter drum during the experiment. At $t = -5$ minutes the chip has been in DI-water for three weeks (first panel in Fig. \ref{fig:afmdefl}a). Flushing with a 1.6 g/L sucrose solution ($\Pi = 0.1$ atm) starts at $t = -2$ min and ends at $t= 0$ min. During the flushing the AFM tip remains in contact with the substrate, but the scans taken during this process are omitted due to the large mechanical disturbance. The first scan after the flush ends at $t= 3$ mins (second panel in Fig. \ref{fig:afmdefl}a). In the first two panels of Fig. \ref{fig:afmdefl}a it is shown that the difference between the scans before (t=-5 min) and after (t=3 min) admitting sucrose is small. However, as time progresses the membrane steadily deflects downwards. The presence of red and yellow regions in the last panels in Fig. \ref{fig:afmdefl}a indicate that the deflection is not uniform, suggesting that the tension distribution in this membrane is not uniform over the membrane area, similar to what was found in other works \cite{davidovikj2016visualizing}.

The height maps in Fig. \ref{fig:afmdefl}a are used to calculate the average deflection of the drum over time as shown in Fig. \ref{fig:afmdefl}c. All height maps are corrected for tilt using the silicon dioxide substrate next to the drum as a reference, whose height is set to zero. Only the part of the graphene, enclosed by the sidewalls (height < 10 nm in Fig. \ref{fig:afmdefl}b), is used to determine the average deflection of the suspended drum. Even in the first scan, before flushing the liquid cell, the membrane height is already lower than that of the substrate. This is due to sidewall adhesion at the edge of the drum as can be seen in the horizontal line cuts in Fig. \ref{fig:afmdefl}b. Due to the sidewall adhesion the membrane has an average height of $-21$ nm in Fig. \ref{fig:afmdefl}c at $t=0$ s. As time progresses, the membrane's average height decreases which will be further analyzed below. 

\begin{figure}[t!]
\includegraphics{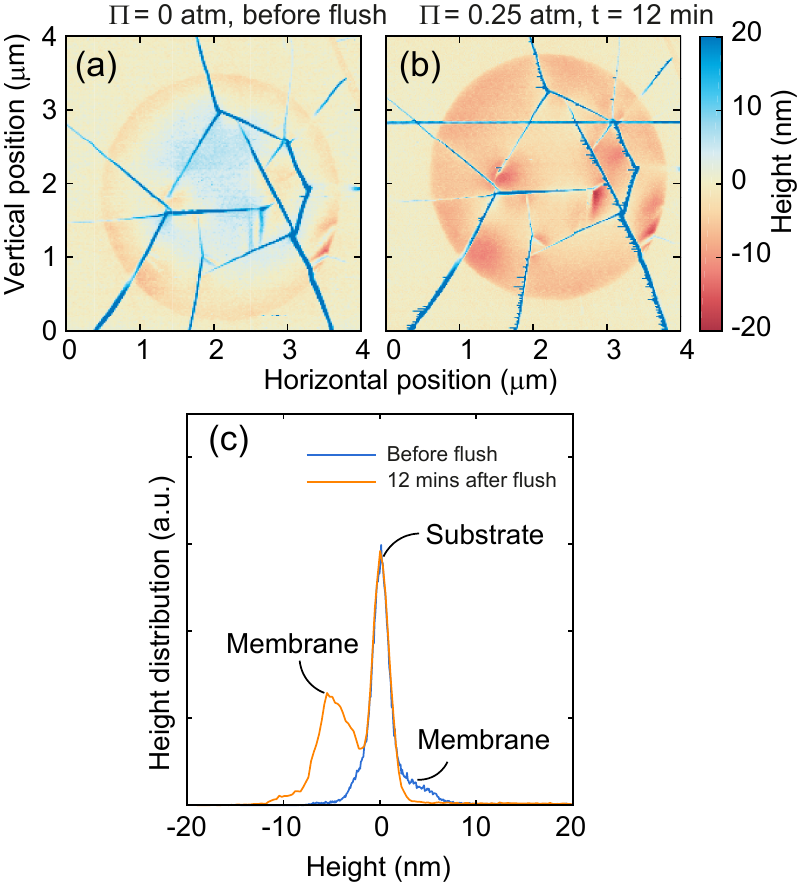}
\caption{Experimental result on a different drum than the one in Fig. \ref{fig:afmdefl}. (a) Height scan of the drum before flushing (the colorbar is the same as (b)). (b) Height scan 12 minutes after flushing the liquid cell with a 4 g/L sugar solution ($\Pi = 0.25$ atm). (c) Height distribution before and after flushing. \label{fig:second}}
\end{figure}
Figure \ref{fig:second} shows the experimental results on a second 3.4 $\mu$m diameter drum. This drum shows clear wrinkles over the surface of the drum, which remain after flushing. Initially, the center of the membrane is higher than the substrate surface (Fig. \ref{fig:second}a), and 12 minutes after flushing with a 4 g/L ($\Pi = 0.25$ atm) sucrose solution the membrane is deflected downward  (Fig. \ref{fig:second}b). This can also be deduced from the change in the height distribution in Fig. \ref{fig:second}c. The downwards deflection can be attributed to the osmotic pressure, consistent with the observations on the first drum in Fig. \ref{fig:afmdefl}. Time-dependent deflection was not studied in this drum, because the AFM lost contact with the substrate during the flushing. Figure \ref{fig:second}b shows the first scan after re-approaching the surface. No significant change in deflection was detected in subsequent scans after $t=12$ min, indicating that this drum permeates significantly faster than the drum in Fig. \ref{fig:afmdefl} and already reached equilibrium between the osmotic and membrane forces during the first scan after flushing.

\section{Model for the time-dependent deflection}
 \begin{figure}[t!]
 \centering
 \includegraphics{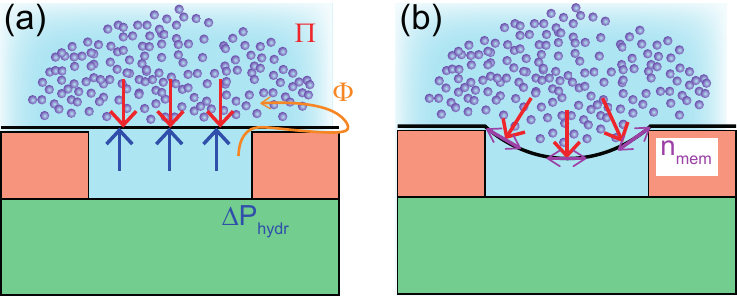}
 \caption{Illustration of the forces on the membrane: (a) immediately after increasing the sucrose concentration in the environment and (b) in thermodynamic equilibrium. \label{fig:osmoforces}}
 \end{figure}
In this section a model is derived to describe the dynamics of the membrane under osmotic pressure. We consider the system to initially be in an equilibrium state, since it is kept in the water for a long time. Then, after introducing the sucrose solution, an osmotic pressure difference develops between the cavity and the environment.
As an estimate of the upper limit, we consider the most extreme case where the graphene-sealed cavity is impermeable to the sucrose-particles and permeable for the water particles. In this case, the concentration of sucrose inside the cavity remains zero and the osmotic pressure difference $\Pi$ is given by van `t Hoff's law \cite{kramer2012five}:
\begin{equation} \label{eq:osmopressure}
\Pi = k_B T c_s,
\end{equation}
where $k_B$ is Boltzmann constant, $T$ is temperature and $c_s$ the concentration of sucrose in the liquid cell environment.

Figure \ref{fig:osmoforces}a schematically shows the forces on the membrane immediately after the environment is flushed with a sucrose solution. The solute (sucrose molecules) generate an osmotic pressure on the membrane pushing it downwards. However, the water in the cavity is incompressible, causing a compressive hydraulic pressure in the cavity  equal to the osmotic pressure. Since the fluid in the environment is at ambient pressure, there is a difference in pressure $\Delta P_{\mathrm{hydr}} = \Pi - P_{\mathrm{mem}}$ between the water inside the cavity and outside the cavity, where $P_{\mathrm{mem}}$ is the pressure resulting from the deflection of the membrane. This hydraulic pressure difference drives water molecules out of the cavity through the pores toward the environment. Using Darcy's law we can express the volumetric flux of water $\Phi$ as:
\begin{equation}\label{eq:flux}
\Phi = \mathfrak{P} (\Pi -   P_{\mathrm{mem}})
\end{equation}
where $ \mathfrak{P}$ is the water permeation rate per unit pressure difference, in units of m$^3$ Pa$^{-1}$ s$^{-1}$.
This flux reduces the volume of the cavity, causing the membrane to deflect downwards thereby tensioning the membrane, resulting in a pressure exerted by the membrane $P_{\mathrm{mem}}$ \cite{bunch2008impermeable}:
\begin{equation}\label{eq:membranepressure}
P_{\mathrm{mem}} = \frac{12 n_{\mathrm{mem}} \delta}{R^2} + \frac{72 E t_g  \delta^3}{3 R^4 (1-\nu)},
\end{equation}
where $n_{\mathrm{mem}}$ is the pre-tension in the membrane, $E$ the Young's modulus, $t_g$ the membrane thickness, $\nu$ Poisson's ratio and $R$ is the radius of the drum. $\delta$ is the average deflection of the drum, defined as: $\delta = h_{\mathrm{avg}} - h_{\mathrm{start}}$, where $h_{\mathrm{avg}}$ the average height of the drum from the AFM measurement and $h_{\mathrm{start}}$ is the average height of the drum in equilibrium when $\Pi = 0$. For simplicity, we assume the deflections of the membrane are sufficiently small to ignore the term proportional to $\delta^3$ in Eq. \ref{eq:membranepressure}. Using Eq. \ref{eq:flux}, the time-dependence of the membrane's average deflection can be obtained by expressing the rate of change in volume of the cavity as: $\Phi = \pi R^2 \mathrm{d}\delta / \mathrm{d} t$, which results in:
\begin{equation}
\frac{\mathrm{d}\delta}{\mathrm{d} t} = \frac{ \mathfrak{P} }{\pi R^2} \Pi -  \frac{ \mathfrak{P} }{\pi R^2} \frac{12 n_{\mathrm{mem}} \delta}{R^2}.
\end{equation}
Solving this differential equation yields the following expression of the time-dependent deflection:
\begin{equation}\label{eq:timedep}
\begin{split}
\delta(t) = - \frac{\Pi R^2}{12 n_{\mathrm{mem}}}\left( 1 - \mathrm{e}^{ - \frac{12 n_{\mathrm{mem}}  \mathfrak{P} }{\pi R^4}t} \right).
\end{split}
\end{equation}

The total deflection ($\delta_{\mathrm{tot}} = |h_{\mathrm{start}} - h_{\mathrm{end}}|$, where $h_{\mathrm{end}}$ is the membrane's final position) the membrane has after reaching thermodynamic equilibrium,
\begin{equation} \label{eq:defl} 
\delta_{\mathrm{tot}} =\frac{ \Pi R^2}{12 n_{\mathrm{mem}}}, 
\end{equation} 
is governed by the tension in the membrane, but does not depend on the permeation rate, since it is obtained when $\Pi = P_{\mathrm{mem}}$. It is observed from Eq. \ref{eq:defl} that a low pre-tension and large radius naturally leads to a larger deflection of the membrane. The characteristic time constant $\tau$ of the time-dependent deflection is governed by both the water permeation rate of and the tension in the membrane:
\begin{equation}\label{eq:tau}
\tau = \frac{\pi R^4}{12 n_{\mathrm{mem}}  \mathfrak{P} }.
\end{equation}
This shows that a small tension leads to a large response time $\tau$ of the system, as more volume needs to be displaced in order to reach thermodynamic equilibrium. As expected, a large permeation rate will lead to a shorter $\tau$. Interestingly, for a constant permeation rate $\mathfrak{P}$, the timeconstant $\tau$ depends strongly on the radius of the membrane $\tau \propto R^4$. Finally, for the extraction of the permeation rate $\mathfrak{P}$ we note that $\tau$ is independent of $\Pi$ according to Eq. \ref{eq:tau}, and is therefore independent of the sucrose concentration used during the experiment. 

\section{Extracting tension and permeation rate}
 By fitting the model in Eq. \ref{eq:timedep} to the experimental data in Fig. \ref{fig:afmdefl}c we extract the pre-tension $n_{\mathrm{mem}}$ and the permeation rate $ \mathfrak{P} $. From the measured total deflection $\delta_{\mathrm{tot}} = 19$ nm we find a pretension $n_{\mathrm{mem}} = 0.13$ N/m. This value is lower than the range found on similar CVD graphene drums based on estimates from their thermal time constant \cite{dolleman2020nonequilibrium,dolleman2020phonon}, but is reasonable compared to other works \cite{lee2008measurement,lopez2015increasing}. Using this value of the pre-tension and deflection, and values for the elastic properties of graphene in literature \cite{lee2008measurement}, the third order term in Eq. \ref{eq:membranepressure} contributes approximately 10\% to the total membrane pressure $P_{\mathrm{mem}}$ in thermodynamic equilibrium. This justifies ignoring the third order term in our analysis. 

The time constant extracted from the fit in Fig. \ref{fig:afmdefl}c is $\tau = \num{1.8e3}$ s. Using Eq. \ref{eq:tau} and the pretension extracted above the water permeation rate is $ \mathfrak{P}  = \num{1.1e-26}$ m$^3$ Pa$^{-1}$ s$^{-1}$. In other works,  the permeation rate of a single pore in graphene $ \mathfrak{P}  = \num{7e-26}$ m$^3$ Pa$^{-1}$ s$^{-1}$ is estimated from experiments \cite{surwade2015water}, while theory \cite{cohen2014water} predicts a value of $ \mathfrak{P}  = \num{1.7e-26}$ m$^3$ Pa$^{-1}$ s$^{-1}$. 
Since our measured value of the permeation rate is similar to the expected value of a single nanopore, we conclude that the drum in Fig. \ref{fig:afmdefl} has a low defect density. Moreover, it is likely that the graphene-silicon dioxide interface is an important pathway for the diffusion of water molecules, similar to the case of gas permeation through graphene nanodrums \cite{lee2019sealing,manzanares2018improved}. The permeation rate per unit area $\mathfrak{P} / \pi R^2 = \num{1.2e-15}$ m Pa$^{-1}$ s$^{-1}$, is 7 orders of magnitude lower than expected for nanoporous graphene \cite{cohen2012water,boretti2018outlook}, supporting the notion of the low defect density in the suspended graphene.

For the second drum in Fig. \ref{fig:second}, we estimate from Fig. \ref{fig:second}(c) an average deflection of approximately 8 nm, resulting in $n_{\mathrm{mem}} = 0.75$ N/m. However, the wrinkles on the membrane are as high as 20 nm, and this additional moment of inertia may cause them to acts as beams that contribute significantly to the overall stiffness of the membrane. Since no significant movement of the membrane is observed after the initial scan at 12 mins, we conclude that this drum was in full thermodynamic equilibrium within 12 minutes. Therefore, the time dependent permation could not be studied in the same detail as the first drum in Fig. \ref{fig:afmdefl}. The downward deflection of the drum, however, does indicate that it is semi-permeable to the sucrose.

\section{Discussion}
Via AFM measurements we have presented evidence that a sucrose concentration gradient across a graphene membrane can cause it to deflect. This effect is attributed to the semipermeability of the graphene membrane for water and sucrose molecules, which causes an osmotic pressure to develop across it. From the total deflection the pre-tension of the membrane can be extracted, while the time-dependence of the deflection is used to extract the water permeation rate of a graphene drum. The value of the permeation rate is close to the value expected for a single pore in the graphene sheet, suggesting the defect density in the suspended CVD graphene is low. 
 Different permeation mechanisms can cause water molecules, that have a kinetic diameter of 0.2 nm, to permeate into the cavity. The first possibility is through the small intrinsic defects in the suspended CVD-graphene sheet, the second is through the wrinkles of the graphene on top of the silicon dioxide (that might act as a lateral channel) and the third is diffusion along the SiO$_2$-graphene interface. Dissolved sucrose molecules have a molecular diameter of 0.9 nm,\cite{price2016sucrose} significantly larger than that of water molecules (0.26 nm), causing it to diffuse at much slower rates (or not diffuse at all) through these permeation pathways. As a consequence, it is expected that the graphene is more permeable to water molecules than to sucrose molecules. 

We note that if the graphene would be permeable for the sucrose, it would be expected that after the initial 'fast' downward deflection of the membrane, it would be followed by a slower upward deflection of the membrane until the sucrose concentration on both sides of the membrane would equilibrate and the membrane would return to its initial position \cite{dolleman2016graphene}. Since no such equilibration or upward deflection was observed within the 60 minutes of measurement, we can conclude that our assumption that the graphene drum is impermeable for sucrose molecules is correct within the experimentally relevant times.

In future work, more information on the permeation mechanism can be revealed by measuring the time constant $\tau$ as a function of drum diameter or membrane thickness, similar to approaches taken to measure the gas permeability of graphene \cite{bunch2008impermeable}. Furthermore, sealing the graphene at the edges will remove the permeation pathways through the silicon dioxide-graphene interface and the wrinkles, allowing permeation to occur only through defects in the graphene membrane \cite{lee2019sealing}. Since our measured value of the permeation rate is close to the expected value of a single nanopore in graphene, such a sealed nanodrum might be suitable to study the (semi-)permeability of individual nanopores. While the AFM method is suitable to study drums with very low permeation rates, introducing many pores or defects in the suspended membrane will significantly increase the permeation rate and therefore decrease $\tau$. This means that faster detection techniques may be more suitable to measure the time-dependent deflection of nanoporous graphene, which could be achieved  by high-speed AFM \cite{katan2011high}; or optical techniques such as colorimetry \cite{cartamil2016colorimetry} or laser interferometry \cite{dolleman2017amplitude}. The main advantage of using AFM  is the possibility to image the inhomogeneities due to wrinkles and nonuniform tension and to study their impact on the deflection profile in detail, which is difficult to achieve with optical techniques. 

The observed deflection of graphene by osmotic pressures is interesting for applications as an osmotic pressure sensor that detects the concentration of solutes, since the deflection of the membrane is a function of the concentration (Eq. \ref{eq:defl}). Graphene-based osmotic pressure sensors use less area than sensors that have been demonstrated using MEMS technology \cite{ch2015design}. Moreover, graphene has excellent chemical stability \cite{EFTEKHARI20171}, large mechanical strength and flexibility; and the small membrane areas achievable with these systems allow for the measurement of extremely low permeation rates while maintaining relatively short response times $\tau$. Reliable fabrication of suspended graphene membranes submerged in water, with tight tolerances on wrinkles and defect densities, is the main challenge for further research in this direction.

\section{Conclusion}
In conclusion, we present evidence that graphene membranes that seal a cavity are deflected by osmotic pressure induced by a sucrose solution. This deflection is characterized by atomic force microscopy in water, demonstrating the feasibility of mechanically characterizing suspended graphene in a liquid environment. When the concentration of sucrose in the surroundings is increased, the membrane deflects downward with an exponential time-dependence which can be explained by our theoretical model. This allows the extraction of the water permeation rate of a single 3.4 $\mu$m diameter drum, which is found to be $\num{1.1e-26}$ m$^3$ Pa$^{-1}$ s$^{-1}$. Since this is close to the expected value of the permeation rate of a single nanopore in graphene, this suggests a low defect density of the suspended graphene sheet. The work opens avenues for studying the semipermeability of graphene membranes via the effect of osmotic pressure on its deflection, and can thus contribute to realizing graphene membrane technology for separation, sensing, energy storage and energy generation applications.

~\\

\section*{Acknowledgements}
The authors thank Applied Nanolayers BV for the supply and transfer of single-layer CVD graphene. This work is part of the research programme Integrated Graphene Pressure Sensors (IGPS) with project number 13307 which is financed by the Netherlands Organisation for Scientific Research (NWO).
The research leading to these results also received funding from the European Union's Horizon 2020 research and innovation programme under grant agreement No 785219 and 881603.

\end{document}